\documentclass[letterpaper,twocolumn,prl,aps,showpacs,superscriptaddress,
floatfix]{revtex4-1}
\usepackage[latin1]{inputenc}
\usepackage{bm}
\usepackage[usenames]{color}
\usepackage{multirow}
\usepackage{amssymb}
\usepackage{amsbsy}
\usepackage{amsmath}
\usepackage{stmaryrd}
\usepackage{graphicx}
\usepackage{epsfig}
\usepackage{placeins}
\usepackage{ulem}
\makeatletter

\newcommand{\deff}{d_\text{eff}}

\begin{document}

\title{Pushing the Limits of the Eigenstate Thermalization Hypothesis\\
towards Mesoscopic Quantum Systems}

\author{R. Steinigeweg}
\email{r.steinigeweg@tu-bs.de}
\affiliation{Institute for Theoretical Physics,
Technical University Braunschweig,
D-38106 Braunschweig, Germany}

\author{A. Khodja}
\affiliation{Department of Physics,
University of Osnabr\"uck,
D-49069 Osnabr\"uck, Germany}

\author{H. Niemeyer}
\affiliation{Department of Physics,
University of Osnabr\"uck,
D-49069 Osnabr\"uck, Germany}

\author{C. Gogolin}
\affiliation{Dahlem Center for Complex Quantum Systems,
Freie Universit\"at Berlin,
D-14195 Berlin, Germany}

\author{J. Gemmer}
\email{jgemmer@uos.de}
\affiliation{Department of Physics,
University of Osnabr\"uck,
D-49069 Osnabr\"uck, Germany}

\date{\today}

\begin{abstract}
In the ongoing discussion on thermalization in closed quantum
many-body systems, the eigenstate thermalization hypothesis (ETH)
has recently been proposed as a universal concept which attracted
considerable attention. So far this concept is, as the name
states, hypothetical. The majority of attempts to overcome
this hypothetical character is based on exact diagonalization
which implies for, e.g., spin systems a limitation to roughly
$15$ spins. In this Letter we present an approach which pushes
this limit up to system sizes of roughly 35 spins, thereby going
significantly beyond what is possible with exact diagonalization.
A concrete application to a Heisenberg spin-ladder which yields
conclusive results is demonstrated. 
\end{abstract}



\maketitle

{\it Introduction}. Due to experiments in ultracold atomic gases
\cite{trotzky2007, hofferberth2007, bloch2008, cheneau2012,
langen2013}, the question of thermalization in closed quantum
systems has experienced an upsurge of interest in recent years, and
the eigenstate thermalization hypothesis (ETH) has become a
cornerstone of the theoretical understanding of thermalizing quantum
many-body systems. The ETH roughly postulates the following
\cite{deutsch1991, srednicki1994, rigol2008}: Eigenstates of a
Hamiltonian $H$ in certain energy regions exhibit properties similar
or equal to the properties of a statistical ensemble, e.g.,
canonical or microcanonical corresponding to that energy region. The
properties in this context can be manifold: expectation values of
certain observables, entropies or purities of  subsystems of an
interacting system, etc. Regardless of its significance in the
debate on thermalization \cite{rigol2008, rigol2012}, it is a
challenging task to ``check'' numerically whether or not the ETH
applies to a specific system and property. If this is done in a
straightforward manner, it requires the diagonalization of the
Hamiltonian \cite{rigol2009, santos2010, rigol2010, steinigeweg2013,
beugeling2013}. Due to the exponential scaling of the Hilbert space
dimension, this is only feasible for rather limited system sizes.
Considering, e.g., spin systems without any symmetries, numerical
diagonalization using state-of-the-art computers and routines is
feasible up to about 15 spins.

\begin{figure}[t]
\includegraphics[width=0.75\columnwidth]{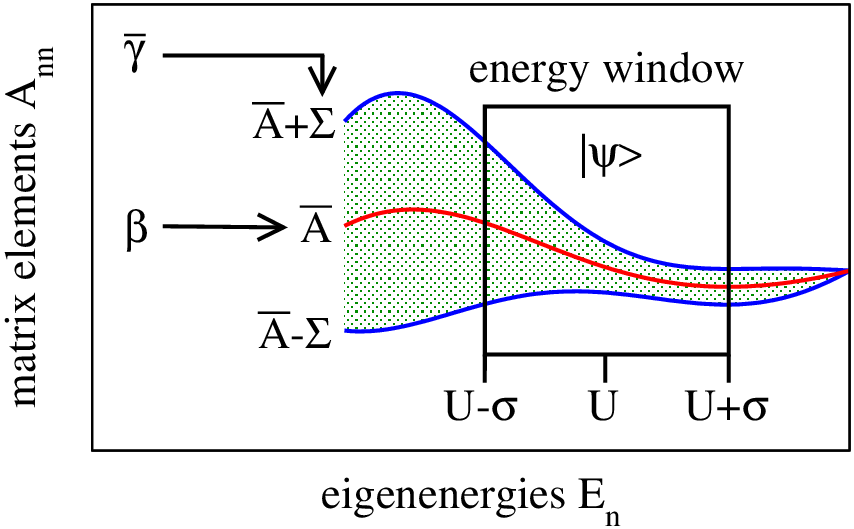}
\caption{(color online) Sketch of the question: In the energy
eigenbasis the diagonal elements $A_{nn} = \langle n | A | n
\rangle$ of a given observable $A$ are in general not a smooth
function of energy, but distributed around their average $\bar{A}$
in a region of width $2 \, \Sigma$; the ETH breaks down when
$\Sigma$ is significantly larger than zero. We present a scheme that
accurately approximates $\bar{A}$ and $\Sigma$ by two other
quantities $\beta$ and $\bar{\gamma}$. The latter can be calculated
on the basis of random state vectors $| \psi \rangle$ that live in
energy windows $[U-\sigma, U+\sigma]$, once operator exponentials
can be applied to pure state vectors in a numerical way.}
\label{Fig1}
\end{figure}

While numerical diagonalization of quantum systems is costly, the
approximation of exponentials of (functions of) the Hamiltonian $H$
applied to a pure state vector, i.e., expressions of the form
\begin{equation}
|\psi (\tau) \rangle \equiv e^{\tau H} |\psi \rangle
\label{exppure}
\end{equation}
($\tau$ being some complex number), has lately seen substantial
progress. Methods in this direction include or are related to
time-dependent density matrix renormalization group (tDMRG)
\cite{vidal2004, daley2004, white2004}, Lanczos \cite{prelovsek2011}
or Chebyshev \cite{deraedt2006} integrator codes. tDMRG allows to
reach system sizes of the order of $100$ or $200$ lattice sites
\cite{karrasch2012}, not only in spin systems \cite{biroli2010}. It
is however somewhat limited regarding system geometries and initial
states: systems must be more or less linear and initial states
usually need to be in some sense close to the ground state. Lanczos
and Chebyshev are at present limited to systems comprising about $35$
spins \cite{prelovsek2011, jin2010, deraedt2007}, however, pure
initial states may be chosen arbitrarily and the only requirement on
the Hamiltonian is a sparse structure when represented w.r.t.~some
reasonable, practically accessible basis \cite{elsayed2013}. This
wider range of applicability makes especially the Chebyshev
integrator a good candidate for future applications of the methods
introduced here.

In the remainder of this Letter we present a scheme that allows for
the computation of ETH-related data on the basis of numerical codes
performing the application of matrix exponentials in the sense of
Eq.~(\ref{exppure}). Apart from the computation of matrix exponentials,
the scheme only requires the ``equilibration'' of the addressed observable
in the sense discussed in Refs.~\cite{reimann2010, linden2009, short2012,
reimann2012} on a time scale within the reach of the matrix-exponentiation
code.

{\it The scheme}. Before explaining the scheme in detail, we specify
more precisely what it eventually provides. Given
a non-degenerate Hamiltonian $H$ and an observable $A$ of a system
with Hilbert space dimension $d$. Then, in the context of the ETH,
the following two quantities are of interest:
\begin{equation}
\bar{A} \equiv \sum_{n=1}^{d} p_n  \langle n | A | n \rangle \, ,
\quad \Sigma^2 \equiv \sum_{n=1}^{d} p_n \langle n | A | n
\rangle^2 - \bar{A}^2
\end{equation}
with $H \, | n\rangle = E_n \, | n\rangle$, $p_n \propto e^{- (E_n
-U)^2/2 \sigma^2}$, and $\sum_n p_n = 1$, i.e., $| n\rangle$ are
eigenstates of the Hamiltonian and $(p_n)_d$ is a Gaussian
probability vector with standard deviation $\sigma$. Thus, $\bar{A}
= \bar{A}(U, \sigma)$ is a weighted average of the expectation
values of $A$ in the eigenstates of $H$ that is most sensitive to an
energy region of width $\sigma$ around $U$, and $\Sigma^2
= \Sigma^2 (U,\sigma)$ is a weighted variance corresponding to this
energy region.

The relation to the ETH is the following: if the ETH applies, the
expectation values are supposed to be a smooth function  of energy,
i.e., the variance $\Sigma^2$ should become small for sufficiently
small $\sigma$. In this sense, $\Sigma^2$ encodes information about
the ETH. The scheme we are going to present in the following allows
for a feasible computation of both, $\Sigma^2$ and $\bar{A}$. The
smaller $\sigma$ is, the more costly this calculation will be.
However, we will present a concrete example in order to demonstrate
the power of our approach.

The computational scheme we present relies on random state vectors.
The fact that few random states suffice to obtain ``non-random''
information on the ETH is closely related to the concept of ``typicality''
\cite{goldstein2006, popescu2006, reimann2007, linden2009, bartsch2009,
sugiura2012, sugiura2013, elsayed2013}. It has been shown that state
vectors drawn at random according to the distribution which is invariant
under all unitary transformations (Haar measure) feature very similar
expectation values for a given observable with high probability
\cite{gemmer2004, popescu2006}. Concretely, using the Hilbert space
average method \cite{gemmer2004}, one finds that the ``Hilbert space
average'' (HA), i.e., the average of the expectation values for an
observable $A$ w.r.t.~to the above distribution and the corresponding
``Hilbert space variance'' (HV) are given by
\begin{eqnarray}
\text{HA}(\langle \psi|A|\psi\rangle) &=& \frac{\text{Tr}(A)}{d} \,
, \label{ham} \\
\text{HV}(\langle \psi|A|\psi\rangle) &=& \frac{1}{d+1} \Big[ \frac{
\text{Tr}(A^2)}{d} - \text{HA}^2 \Big ] \, .
\label{hv}
\end{eqnarray}
Equipped with these results, we now describe the scheme.

{\it First step.} We start by defining an operator $C$,
\begin{equation}
C \equiv e^{-\frac{(H-U)^2}{4\sigma^2}} \, ,
\label{eshiftop}
\end{equation}
which we will later use as an ``energy filter''. (Similar ``filters''
have been used previously, e.g., in Refs.~\cite{presilla1995, Garnerone13a,
Garnerone13b}). We are interested in computing $\text{Tr}(C^2)$ since this
will be needed as a normalization constant below. To this end we consider
the random variable
\begin{equation}
\alpha \equiv d \, \langle \psi| C^2 |\psi \rangle \, ,
\label{alphadef}
\end{equation}
where $|\psi \rangle$ is a random state vector drawn according to the
above distribution. If we are able to apply matrix exponentials to random
pure state vectors, we are able to compute random realizations of $\alpha$.
Using Eq.~(\ref{ham}), we immediately find
\begin{equation}
\text{HA} (\alpha) = \text{Tr}(C^2) = \sum_{n=1}^{d} e^{-\frac{(E_n
-U)^2}{2\sigma^2} } \, . \label{hamalpha}
\end{equation}
The average of $\alpha$ is the quantity we are interested in. If the
distribution of $\alpha$ was broad, estimating its HA would be costly
since it would require computing many realizations of $\alpha$. But
from Eq.~(\ref{hv}) we may directly read off an upper bound on the
variance of $\alpha$:
\begin{equation}
\text{HV} (\alpha) < \frac{d}{d+1} \text{Tr} (C^4) \leq \sum_{n=1}^d
e^{-\frac{(E_n-U)^2}{\sigma^2} } \label{hvalpha}
\end{equation}
We cannot directly compute the sums in Eqs.~(\ref{hamalpha}) and
(\ref{hvalpha}); however, we can reasonably guess their scaling with
the density of states  $n(E)$. If $n(E)$ is a sufficiently smooth
function to be linearized on a scale of $\sigma$ around $U$, then
the sums yield approximately
\begin{equation}
\text{Tr}(C^2) \approx \sqrt{2\pi} \sigma \, n(U) \, , \quad
\text{Tr}(C^4) \approx \sqrt{\pi} \sigma \, n(U) \, .
\label{trc2simp}
\end{equation}
Let us abbreviate $\sigma \, n(U) \equiv \deff$. The meaning of
$\deff$ is that of an effective dimension. The number of states in
the respective energy window is roughly $\deff$. If the size of a
quantum system is increased while $\sigma$ is kept fixed, $\deff$
can be expected to become very large rather quickly. Thus,
w.r.t.~$\deff$, the average of $\alpha$ and an upper bound to its
variance read in the limit of large $d$
\begin{equation}
\text{HA} (\alpha) \approx \sqrt{2\pi} \, \deff \, , \quad \text{HV}
(\alpha) \lesssim \sqrt{\pi} \, \deff \, .
\end{equation}
Since the standard deviation scales with the square root of the
variance, the distribution of $\alpha$ with mean $\text{HA} (\alpha)
\propto \deff$ has the width $\sqrt{\text{HV} (\alpha)} \propto
\sqrt{\deff}$. Because the mean of $\alpha$ is the quantity of
interest, calculating one $\alpha$ from one random $|\psi \rangle$
amounts to the determination of the wanted quantity with a relative
error on the order of $1/\sqrt{\deff}$. Thus, if $\deff$ is large
enough, calculating only very few realizations of $\alpha$ will
suffice.

\begin{figure}[t]
\includegraphics[width=0.90\columnwidth]{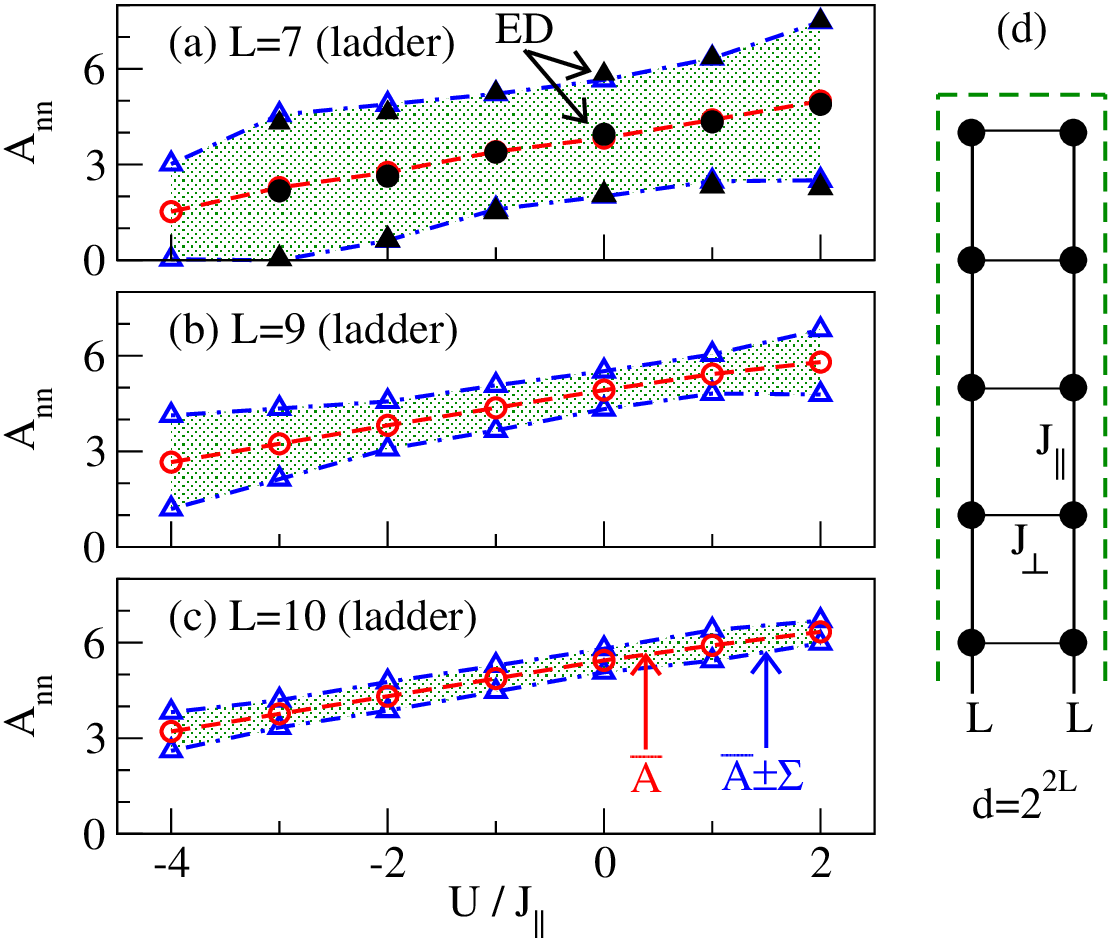}
\caption{(color online) The ``cloud'' of matrix elements $A_{nn}$ for
(a) $L=7$, (b) $9$, and (c) $10$, obtained numerically using the scheme
and $10$ random state vectors. Panel (a) further provides a comparison
with exact-diagonalization (filled symbols), clearly indicating negligibly
small deviations of our scheme already for $L=7$. Panel (d) is a sketch
of the ladder model (solid lines) and its reduction to a chain (dashed
lines).}
\label{Fig2}
\end{figure}

{\it Second step.} Next we define
\begin{equation}
\rho \equiv \frac{C^2}{\text{Tr}(C^2)} \, , \label{shiftopnorm}
\end{equation}
which is a positive operator with trace one, i.e., a quantum state,
and consider its application to a pure state vector. We note that
$\langle m | \rho | n \rangle = p_n \, \delta_{nm}$. In order
to determine $\bar{A}$, we also define and consider
\begin{equation}
\beta \equiv d \langle \psi | \sqrt{\rho} A \sqrt{\rho} | \psi
\rangle \, . \label{typvarcheck}
\end{equation}
Using again Eqs.~(\ref{ham})
and (\ref{hv}), and following the same line of reasoning as in the
context of $\alpha$, we readily find the average of $\beta$ and an
upper bound to its variance:
\begin{eqnarray}
\text{HA}(\beta) &=& \text{Tr} (\rho A) = \bar{A} \, , \\
\text{HV} (\beta) &<& \frac{d}{d+1} \text{Tr} (\rho A \rho A)
\end{eqnarray}
Again the average of $\beta$ is the quantity of interest, and its
computation is feasible if $\text{HV} (\beta)$ is small. (Note
that it is not $\text{HV} (\beta)$ from which the desired $\Sigma$
is eventually calculated.) To upper bound the variance, we write:
\begin{equation}
\text{Tr} (\rho A \rho A) = \sum_{m,n} p_n \langle n |A| m \rangle
p_m \langle m |A| n \rangle
\end{equation}
From Eqs.~(\ref{eshiftop}), (\ref{trc2simp}) [l.h.s.], (\ref{shiftopnorm})
we find $p_n \leq 1/\deff$, which implies:
\begin{equation}
\text{Tr} (\rho A \rho A) \leq \sum_{m,n} \frac{p_n
\langle n |A| m \rangle \langle m | A |n \rangle}{\deff} =
\frac{\text{Tr} (\rho A^2)}{\deff}
\end{equation}
If $A$ is  taken to be traceless (w.l.o.g.), for large $d$ the
variance $\text{HV} (\beta)$ is essentially upper-bounded by a term
that scales as the, say, largest squared eigenvalue of $A$ divided
by the effective dimension $\deff$. The largest eigenvalue of physical
observables scales at most polynomially with system size, the effective
dimension typically increases exponentially. Hence, by calculating only
a few $\beta$, it should be possible to determine $\bar{A}$ within a
relative error $\propto 1/\sqrt{\deff}$.

{\it Third step.} Next, aiming at $\Sigma$, we consider
\begin{equation}
\gamma(t) \equiv d \langle \psi | \sqrt{\rho} A(t) A \sqrt{\rho} |
\psi \rangle \, , \label{defgamma}
\end{equation}
where $A(t)$ refers to the Heisenberg picture. Again, given the possibility
to apply matrix exponentials to arbitrary state vectors, $\gamma(t)$ can be
computed for random state vectors $|\psi\rangle$. To proceed, one has to
require that $\gamma(t)$ not only relaxes with time to some value and then
does not deviate much from that value \cite{reimann2010, linden2009, short2012,
reimann2012}, but, moreover, this must happen on time scales which are
``short'' compared to the time scales over which $\gamma(t)$ can be
approximated numerically. If this applies, a time average from the relaxation
time $t_1$ to the largest time $t_2$ reachable with the given resource will
very accurately approximate the average over infinite time. Whether or not
the above condition holds has to be guessed (or postulated), as well as the
precise choice of $t_1$ and $t_2$. However, the graph of $\gamma(t)$ itself
may give good evidence and suggest a reasonable choice for $t_1$, $t_2$. If
the above holds, we find
\begin{equation}
\overline{\gamma} = \frac{1}{t_2-t_1} \int_{t_1}^{t_2} \!
\mathrm{d}t \, \gamma(t) \approx d \, \langle \psi | \sqrt{\rho} A_D A
\sqrt{\rho} | \psi \rangle  \label{defovgamma}
\end{equation}
with $A_D$ being the diagonal part of $A$ in the energy eigenbasis.
The random variable $\overline{\gamma}$ is very similar to $\beta$,
c.f.~Eq.~(\ref{typvarcheck}). Going through precisely the same arguments
that follow Eq.~(\ref{typvarcheck}), one finds (still for large $d$)
\begin{eqnarray}
\text{HA} (\overline{\gamma}) &\approx& \text{Tr} (\rho A_D A)
= \Sigma^2 + \bar{A}^2 \, , \\
\text{HV}(\overline{\gamma}) &\lesssim& \frac{1}{\deff} \text{Tr}
(\rho A_D A^2 A_D) \, .
\end{eqnarray}
As before, if the largest eigenvalue of $A$ does not scale exponentially
with system size, $\Sigma$ may be determined from a few
realizations of $\overline{\gamma}$ within an error $\propto 1/\sqrt{\deff}$.

\begin{figure}[t]
\includegraphics[width=0.75\columnwidth]{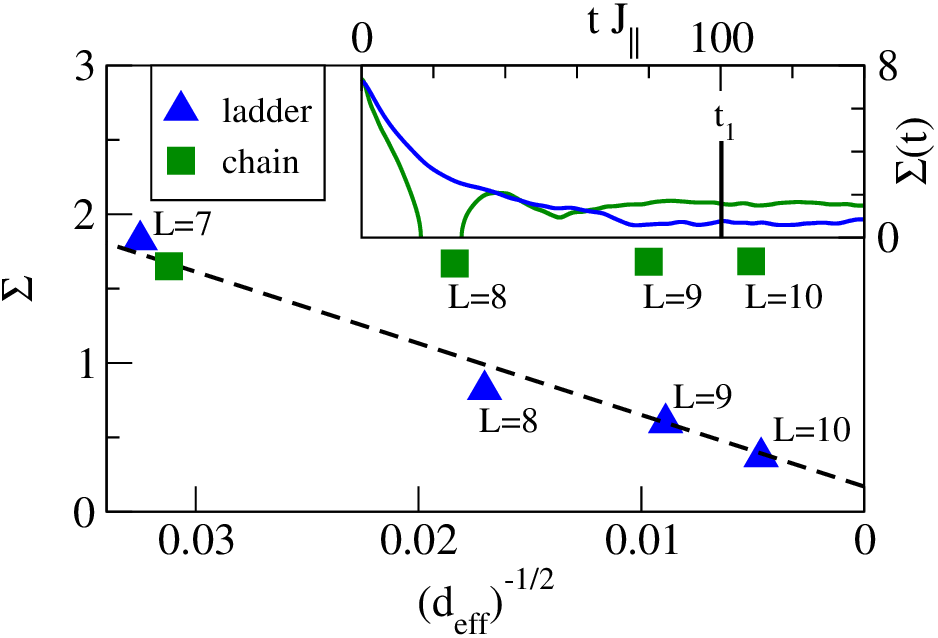}
\caption{(color online) The ``cloud'' width $\Sigma$ as a function of the
effective-dimension power $(\deff)^{-1/2}$ at energy $U=0$ for the non-integrable
ladder in Fig.\ 2 and an integrable chain. As a guide to the eye, the line indicates
a function $\propto (\deff)^{-1/2}$, where the offset at $(\deff)^{-1/2} \to 0$
($L \to \infty$) is given by $\sigma \, \mathrm{d}\bar{A}/\mathrm{d}U \approx 0.17$
as the product of the chosen energy-window width $\sigma$ and the ``cloud'' slope
$\mathrm{d}\bar{A}/\mathrm{d}U$, see the discussion in the text for details. The
inset shows the underlying graph of $\Sigma(t)$ for $L=10$, with
the relaxation time $t_1$ indicated.}
\label{Fig3}
\end{figure}

{\it Application.} Eventually, we  illustrate the introduced scheme
using a Heisenberg spin ladder of length $L$ without periodic boundary
conditions as an example. The Hamiltonian $H = J_\parallel H_\parallel
+ J_\perp H_\perp$ reads ($\hbar=1$)
\begin{eqnarray}
&& H_\parallel = \!\! \sum \limits_{r=1}^{L-1} \! \sum_{i=1}^2
S^x_{r,i} S^x_{r+1,i} + S^y_{r,i} S^y_{r+1,i} + \Delta \, S^z_{r,i}
S^z_{r+1,i} \, , \nonumber \\
&& H_\perp = \sum_{r=1}^{L} S^x_{r,1} S^x_{r,2} + S^y_{r,1}
S^y_{r,2} + \Delta \, S^z_{r,1} S^z_{r,2} \, , \label{model}
\end{eqnarray}
where $S_r^{x,y,z}$ are spin-1/2 operators at site $(r,i)$,
$J_\parallel > 0$ is the antiferromagnetic exchange coupling
constant along the legs, and $J_\perp=0.2\,J_\parallel$ is a small
rung interaction. The exchange anisotropy $\Delta = 0.6$ is chosen
to realize several non-thermalizing properties of the legs alone
\cite{steinigeweg2013}. The Hamiltonian preserves the total magnetization
$S^z_{\text{total}}$ and is non-degenerate except for a two-fold degeneracy
due to ``particle-hole symmetry'' \cite{steinigeweg2013, znidaric2013}. We
choose the largest ``half-filling'' subspace  $S^z_{\text{total}}=0$. This
non-integrable ladder reduces to a integrable chain if $J_\perp=J_\parallel$
and only the first rung is kept in $H_\perp$, see Fig.\ \ref{Fig2} (d).

We study the magnetization difference $\delta M = \sum_{r=1}^{L} S^z_{r,1}
- S^z_{r,2}$ of the two legs. Due to the ``particle-hole symmetry'', any
energy eigenstate must necessarily yield $\langle n | \delta M | n \rangle
= 0$ and describes the magnetization as being equally distributed between
the two legs. Hence, one could say that the ETH is fulfilled w.r.t.~the
observable $\delta M$. This, however, does not mean that in every single
measurement the magnetization on each of the two legs is $\pm L/2$, which
would be true only if $\langle n | \delta M^2 | n \rangle = 0$. But the
latter is not trivially fulfilled since $\delta M^2$  shares ``particle-hole''
symmetry other than $\delta M$. Therefore, $\delta M^2$ may possibly vary
from eigenstate to eigenstate.

A recent publication \cite{niemyer2013} reported that $\langle
\psi(t) | \delta M | \psi(t) \rangle$ and $\langle \psi(t) | \delta
M^2 | \psi(t) \rangle$ relax to some values, almost regardless of
the initial state (restricted to a window of energy). While this is
not very surprising for $\delta M$, it hints towards the validity of
the ETH w.r.t.~$\delta M^2$. Since the behavior described in
Ref.~\cite{niemyer2013} becomes pronounced only above $L=8$,
numerically checking the ETH is a highly non-trivial task. We
show that with our proposed scheme it is however possible
to convincingly verify numerically the validity of the ETH
w.r.t.\ the observable $A=\delta M^2$.

We begin with state vectors $| \psi(0) \rangle = \sum_i c_i |i
\rangle$, where the set of state vectors ${| i \rangle}$ is the
Ising basis in the convenient spin-$\uparrow$/$\downarrow$
representation, in the subspace $S^z_\text{total}=0$. The
coefficients $c_i$ are obtained by generating independent Gaussian
random numbers with mean zero and variance one for the real and
imaginary part. In order to compute realizations of $\alpha$ from
Eq.~(\ref{alphadef}), the ``energy-filter'' operator $C$ in
Eq.~(\ref{eshiftop}) is approximated by a fourth order Runge-Kutta
integrator \cite{rk4}, iterating in imaginary time with a discrete
time step $\delta t$ until the chosen energy window is reached. For
all calculations we choose $\sigma=0.37$, which is small compared
to the width of the spectrum of $H$. Using the same imaginary-time
iteration, the mean of $\beta$ in Eq.~(\ref{typvarcheck}) can also
be approximated. Similarly, a real-time iteration \cite{rk4} provides
the technical foundation for the calculation of $\gamma(t)$ in
Eq.~\eqref{defgamma}, starting from two different energy-filtered
initial state vectors, first $C | \psi(0) \rangle$ and second
$\delta M^2 C | \psi(0) \rangle$. The choice of $t_1$ and $t_2$ in
Eq.~(\ref{defovgamma}) is made manually by reading off times where
$\gamma(t)$ does not show any kind of dynamics apart from minor
oscillations (see the inset of Fig.~\ref{Fig3}).

Figure~\ref{Fig2} (a) compares the scheme to results from exact
diagonalization for $L=7$. Apparently, the agreement is remarkably
good for a rather small system, supporting that the mean of $\beta$
and $\bar{\gamma}$ indeed yield good approximations of the exact
average $\bar{A}$ and variance $\Sigma$. Figure~\ref{Fig2} (b)
shows the results for the ``cloud'' center $\bar{A} \approx \beta$
and the ``cloud'' width $\Sigma \approx \sqrt{\bar{\gamma} - \beta^2}$
but now for $L=9$ and $10$. While $\Sigma$ very clearly decreases with
$L$, $\bar{A} \propto L$ due $\delta M$ being extensive \cite{niemyer2013},
but the slope of $\bar{A}$ as a function of energy stays the same.

The question of how the ``cloud'' width $\Sigma$ scales with the
system size can be answered by plotting it against the effective
dimension $\deff$ in Eq.~(\ref{trc2simp}) for some energy interval
in Fig.~\ref{Fig2}. For convenience, Fig.~\ref{Fig3} shows $\Sigma$
vs.\ the effective-dimension power $(\deff)^{-1/2}$. Clearly,
Fig.~\ref{Fig3} supports the scaling $\Sigma \propto (\deff)^{-1/2}$.
Such a scaling is expected for random Hamiltonians \cite{beugeling2013}
and is consistent with the non-integrability of our model. Even though
there is a remaining offset at $(\deff)^{-1/2} \to 0$ ($L \to \infty$),
this offset does {\it not} indicate the breakdown of the ETH for our
observable and model. In fact, the offset is a minimum ``cloud'' width
$\sigma \, \mathrm{d}\bar{A}/ \mathrm{d}U \approx 0.17$ given by the
product of the chosen energy-window width $\sigma$ and the ``cloud''
slope $\mathrm{d}\bar{A}/\mathrm{d}U$. To illustrate how the breakdown
of the ETH can be detected by our approach, Fig.\ \ref{Fig3} shows also
results on the integrable chain ($J_\perp=J_\parallel$ and a single rung
in $H_\perp$), where $\Sigma$ does not depend on system size.

{\it Conclusion.} In this paper we presented an innovative scheme
for deciding whether or not the ETH is valid in a given closed
many-body quantum system of finite but very large Hilbert space
dimension. Using the framework of typicality, we showed that both,
average and variance of the diagonal matrix elements of a given
observable in the energy eigenbasis can be calculated from a single
or few pure state vectors, if the application of operator exponentials
is available in a numerical way. We demonstrated the latter for a
prototypical spin model by using a Runge-Kutta iterator. While
Runge-Kutta will allow for more than $20$ spins in cases with several
symmetries \cite{steinigeweg2014}, sophisticated algorithms like Chebyshev
integrators, when used with our scheme, will enable almost exact
studies of the ETH in systems of up to $35$ spins \cite{jin2010,
deraedt2007} and provide further insight into thermalization in large
closed quantum many-body systems.

We gratefully acknowledge financial support by the {\it Deutsche
Forschungsgemeinschaft} and the {\it Studienstiftung des Deutschen
Volkes}.

\end{document}